\documentclass[12pt]{iopart}
\usepackage{iopams}
\usepackage[utf8]{inputenc}
\usepackage{graphicx}

\begin{document}

\title[The interface of H$_2$O and GaP(100) studied by PES and RAS]{The interface of GaP(100) and H$_2$O studied by photoemission and reflection anisotropy spectroscopy}

\author{Matthias M May,$^{1,2}$ Oliver Supplie,$^{1,2}$ Christian Höhn,$^1$ Roel van de Krol,$^1$ Hans-Joachim Lewerenz,$^{1,3}$ and Thomas Hannappel$^{1,4,5}$}

\address{$^1$ Helmholtz-Zentrum Berlin für Materialien und Energie, Institute for Solar Fuels, Hahn-Meitner-Platz 1, D-14109 Berlin, Germany}
\address{$^2$ Humboldt-Universität zu Berlin, Institut für Physik, Newtonstr. 15, D-12489 Berlin, Germany}
\address{$^3$ California Institute of Technology, Joint Center for Artificial Photosynthesis, 1200 East California Boulevard, Pasadena, CA 91125, USA}
\address{$^4$ Technische Universität Ilmenau, Institut für Physik, Gustav-Kirchhoff-Str. 5, D-98693 Ilmenau, Germany}
\address{$^5$ CiS Forschungsinstitut für Mikrosensorik und Photovoltaik GmbH, Konrad-Zuse-Str. 14, D-99099 Erfurt, Germany}
\ead{Matthias.May@helmholtz-berlin.de}
\begin{abstract}

We study the initial interaction of adsorbed H$_2$O with P-rich and Ga-rich GaP(100) surfaces. Atomically well defined surfaces are prepared by metal-organic vapour phase epitaxy and transferred contamination-free to ultra-high vacuum, where water is adsorbed at room temperature.  Finally, the surfaces are annealed in vapour phase ambient. During all steps, the impact on the surface properties is monitored with in-situ reflection anisotropy spectroscopy (RAS). Photoelectron spectroscopy and low-energy electron diffraction are applied for further in-system studies. After exposure up to saturation of the RA spectra, the Ga-rich $(2 \times 4)$ surface reconstruction exhibits a sub-monolayer coverage in form of a mixture of molecularly and dissociatively adsorbed water. For the  $p(2 \times 2)/c(4 \times 2)$ P-rich surface reconstruction, a new $c(2 \times 2)$ superstructure forms upon adsorption and the uptake of adsorbate is significantly reduced when compared to the Ga-rich surface. Our findings show that microscopic surface reconstructions of GaP(100) greatly impact the mechanism of initial interface formation with water, which could benefit the design of e.g. photoelectrochemical water splitting devices.

\end{abstract}


\submitto{\NJP}

\maketitle

\section{Introduction}

Harvesting solar energy and storing it in the form of hydrogen can be achieved in a single device, a photoelectrochemical solar cell \cite{Fujishima_photolysis_1972}. In such a device, the light-absorbing  semiconductor is brought in contact with an electrolyte so that the photo-generated electrons or holes can directly reduce or oxidize water, respectively. In such a device, several challenges have to be addressed that do not arise in photovoltaics, such as corrosion and proper energetic alignment to the redox potentials of the aqueous electrolyte \cite{Gerischer_stability_sc_electrodes_1977}. III-V semiconductors are a promising absorber material class for this application due to a high flexibility in opto-electronic properties, with energy gap as well as band alignment tunable via the growth of ternary (quarternary) compounds \cite{ChemPhysChem_III-V_2012, PEC_book_chapter_2013}. Devices however, which are simultaneously efficient and (photo)chemically stable, have yet to be realised.

A microscopic understanding of both morphology and electronic structure  at the solid-liquid interface is essential for the design of the semiconductor surface at this phase boundary. Initial oxide formation induced upon contact with the electrolyte, for example, can either hinder charge-transfer to the electrolyte \cite{Kaiser2012} or favour it due to proper band alignment, when formed in a controlled manner \cite{Lewerenz2010}. The oxidation of InP, a semiconductor closely related to GaP, was for instance found to be very sensitive to the surface reconstruction \cite{Chen_InP_oxidation_RAS_2002}. The question arises to what extent different surface reconstructions impact the interaction between water and the semiconductor, and if this could benefit device designs, as was shown for heteroepitaxial growth of tunnel junctions \cite{Sagol_basic_concepts_multijunction_2007}.

Model experiments in the literature involved water adsorption in ultra-high vacuum (UHV) to gain insight into the surface modifications of semiconductors induced by H$_2$O \cite{Jaegermann_modern_asp_elchem, Schmeisser_comparative_adsorption_Si}, as typical surface science tools such as photoelectron spectroscopy cannot be applied in a fully realistic liquid environment. The initial contact between water and the III-V semiconductors InP and GaP has been subject of several theoretical investigations \cite{Wood_H2O_semoconductor_2010, Wood_GaP_InP_2012, Jeon_GaP_2012}, but experimental data only exists for InP(110) \cite{Montgomery_H2O_InP_1982, Henrion_water_InP_2000} and not for GaP to our knowledge. 

Wood et al. studied chemisorbed oxygen and hydroxyl groups on GaP and InP (100) surfaces with density-functional theory, identifying the most probable structural motifs to be Ga-[OH]-Ga, Ga-OH, Ga-O-P and Ga-O-Ga \cite{Wood_GaP_InP_2012}. The latter is thought to create traps for holes and to initiate corrosion of the semiconductor photocathode under working conditions. For adsorbed hydroxyl on the Ga-rich $(2 \times 4)$ surface reconstruction, Ga-[OH]-Ga bridge configurations and Ga-OH atop configurations are expected to be energetically most favourable, with neighbouring OH groups stabilized further by hydrogen bonding \cite{Wood_GaP_InP_2012}. The stability of the chemisorbed states are expected to depend on the bond topology, which should in principle be visible to surface-sensitive surface science tools. At room temperature, however, the surfaces will probably exhibit not only the energetically most favourable state, but also other states due to dynamical interconversion. Jeon et al. have modelled the interaction of a single H$_2$O molecule with the Ga-rich $(2 \times 4)$ surface reconstruction of GaP(100) \cite{Jeon_GaP_2012}. They found a three-step process to be most likely, where H$_2$O is initially adsorbed in a molecular state, then dissociated into HO/H and finally forms Ga-O-Ga bridges desorbing molecular hydrogen.

\begin{figure}[h]
	\begin{center}
	\includegraphics[width=.65\linewidth]{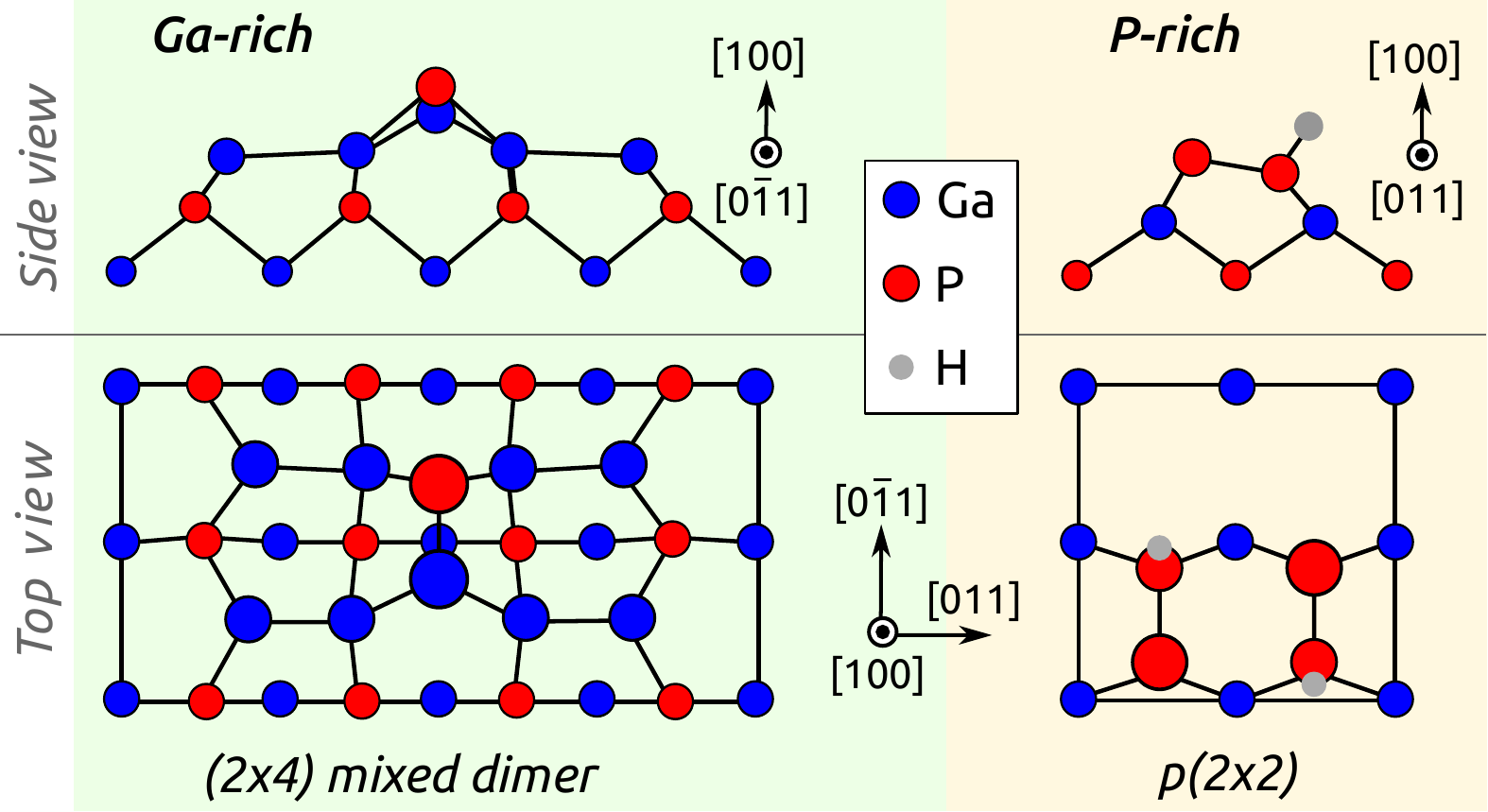}
	\caption{Ball-and-stick model of the considered surface reconstructions of GaP(100). The left side shows the Ga-rich, $(2 \times 4)$ mixed dimer reconstruction, the right side the P-rich  $p(2 \times 2)/c(4 \times 2)$ reconstruction. Blue represents Ga atoms, red P atoms and grey H atoms; size increases for atoms that are closer to the top.}
	\label{fig:surface-reconstructions}
	\end{center}
\end{figure}

The well-defined P-rich, $p(2 \times 2)/c(4 \times 2)$ and the Ga-rich, $(2 \times 4)$ mixed dimer surface reconstructions \cite{Hahn_P-rich_GaP_calculated_2003, Frisch_Ga-rich_GaP_1999}, which are typical for films grown by metal-organic vapour phase epitaxy (MOVPE) in hydrogen ambient, are the initial points of our experiments and juxtaposed in figure \ref{fig:surface-reconstructions}. The Ga-rich surface reconstruction features a mixed Ga-P dimer on top of a layer of Ga atoms. The P-rich surface reconstruction is formed of buckled P-P dimers on top, which are stabilized by one hydrogen atom per dimer. The orientation of the H atom in adjacent rows of P dimers leads to a mixture of $p(2 \times 2)$ and $c(4 \times 2)$ phases, which can inter-convert due to a flipping motion of the P dimers at room temperature \cite{Kleinschmidt_dimer_flipping_2011}.

In this paper, we combine photoelectron spectroscopy (PES) and LEED in model experiments with in-situ reflection anisotropy spectroscopy (RAS) \cite{Aspnes_Studna_RAS_1985} to investigate water-induced modifications of the surface regarding electronic structure and morphology. We find a distinctly different interaction of the P-rich and the Ga-rich surface reconstructions with water, identifying adsorbate-related signatures in RAS and potential reaction mechanisms.

\section{Experimental}

Investigations on the semiconductor-liquid interface in adsorption experiments necessitate well-defined and clean semiconductor surfaces as a point of reference for the subsequent adsorbate-induced surface modifications. Metal-organic vapour phase epitaxy was used here to prepare two different surface reconstructions in hydrogen atmosphere at near-ambient pressure. Optical in-situ growth control was achieved with RAS enabling a clear distinction between the P- and the Ga-rich surface already in the MOVPE reactor (Aixtron AIX 200) during growth and thereby well-defined preparation \cite{Toeben_RAS_LEED_STM_GaP_2001}.

RAS is an optical spectroscopic method sensitive to dielectric anisotropies of for example two perpendicular crystal directions in cubic semiconductors \cite{Aspnes_Studna_RAS_1985}. RAS can be applied in both UHV and liquid environments and has previously been applied to semiconductor-adsorbate interfaces as well as to metal-liquid interfaces \cite{Chen_InP_oxidation_RAS_2002, Witkowski_RAS_adorbates_2006,Smith_RAS_Au_electrolyte_2009}. The sample is irradiated with linearly polarized light at near-normal incidence and the difference $\Delta r$ in reflection along the two axes -- $[0\bar{1}1]$ and $[011]$ for (100) surfaces -- is measured and normalized to the arithmetic mean of the total reflection, $r$:

\begin{equation}
\label{eq:ras}
 \frac{\Delta r}{r}=2\frac{r_{[0\bar{1}1]}-r_{[011]}}{r_{[0\bar{1}1]}+r_{[011]}}  ,  r\in\mathbb{C}
\end{equation}

A commercial spectrometer (LayTec EpiRAS 200) was used for both the measurements in the MOVPE reactor and in the UHV setup, for details, see ref. \cite{Haberland_RAS_setup_1999}.

Initial surface preparation in the MOVPE reactor consisted of deoxidation of a GaP(100) wafer under hydrogen atmosphere, followed by homoepitaxial growth of an about 200\,nm thick GaP buffer layer with the precursors tertiarybutylphosphine (TBP) and triethylgallium. After buffer growth, either the P-rich or the Ga-rich surface reconstruction was prepared by dedicated annealing steps applying optical in-situ control with RAS \cite{Toeben_RAS_LEED_STM_GaP_2001, Letzig_P-H_InP_2005}. The P-rich surface reconstruction was achieved by cooling down the sample after growth to 300$^\circ$C (temperatures were corrected for an offset of approximately 10\,K) under TBP supply and finally annealing it for 10\,min at 410$^\circ$C without TBP, while the Ga-rich surface preparation necessitated an annealing for 5\,min at 700$^\circ$C \cite{Hannappel_InP_GaP_2001}. The subsequent, contamination-free transfer from the MOVPE reactor to the UHV setup employed a dedicated transfer system \cite{Hannappel_Transfer_chamber_2004} with a mobile UHV shuttle and base pressures in the low $10^{-10}$\,mbar range.

The PES system includes a He discharge lamp, a monochromated X-ray source (Specs Focus 500 with monochromated Al K$_\alpha$ and Ag L$_\alpha$ sources) as well as a hemispherical analyser (Specs Phoibos 100). The surface sensitivity of XPS was increased by tilting the samples to create a take-off angle of 60$^\circ$ against normal emission, decreasing the information depth of the photoelectrons via their inelastic mean free path. Furthermore, a LEED system (Specs ErLEED 100-A) and an adsorption chamber with an optical viewport for RAS are attached. Adsorption of ultra-pure water in a dedicated UHV chamber (base pressure lower $10^{-8}$\,mbar) was realized through a leak valve at room temperature and H$_2$O partial pressures in the order of $10^{-5}$\,mbar. Adsorbate dosages were measured in Langmuir (L) employing the uncorrected pressure rise in the chamber. Cleanliness of the water vapour was checked with mass spectrometry. During the exposure, the sample surface could be monitored continuously with RAS allowing in-situ observation of water-induced surface modifications. Afterwards, samples were again inspected in-system with LEED and PES. A transfer back to the MOVPE reactor via the UHV shuttle enabled annealing in ultra-pure hydrogen and nitrogen at $p=100$\,mbar  with RAS in-situ control.

\section{Results}

We will first present our findings for the Ga-rich surface of GaP(100), as its behaviour with respect to H$_2$O exposure has been subject of several theoretical studies \cite{Wood_GaP_InP_2012, Jeon_GaP_2012}. The P-rich surface, which can only be prepared under more specific conditions due to its hydrogen stabilisation, will be the subject of section \ref{sec:P-rich}.

\subsection{Ga-rich, mixed dimer surface reconstruction}

\begin{figure}[h]
	\begin{center}
	\includegraphics[width=.95\linewidth]{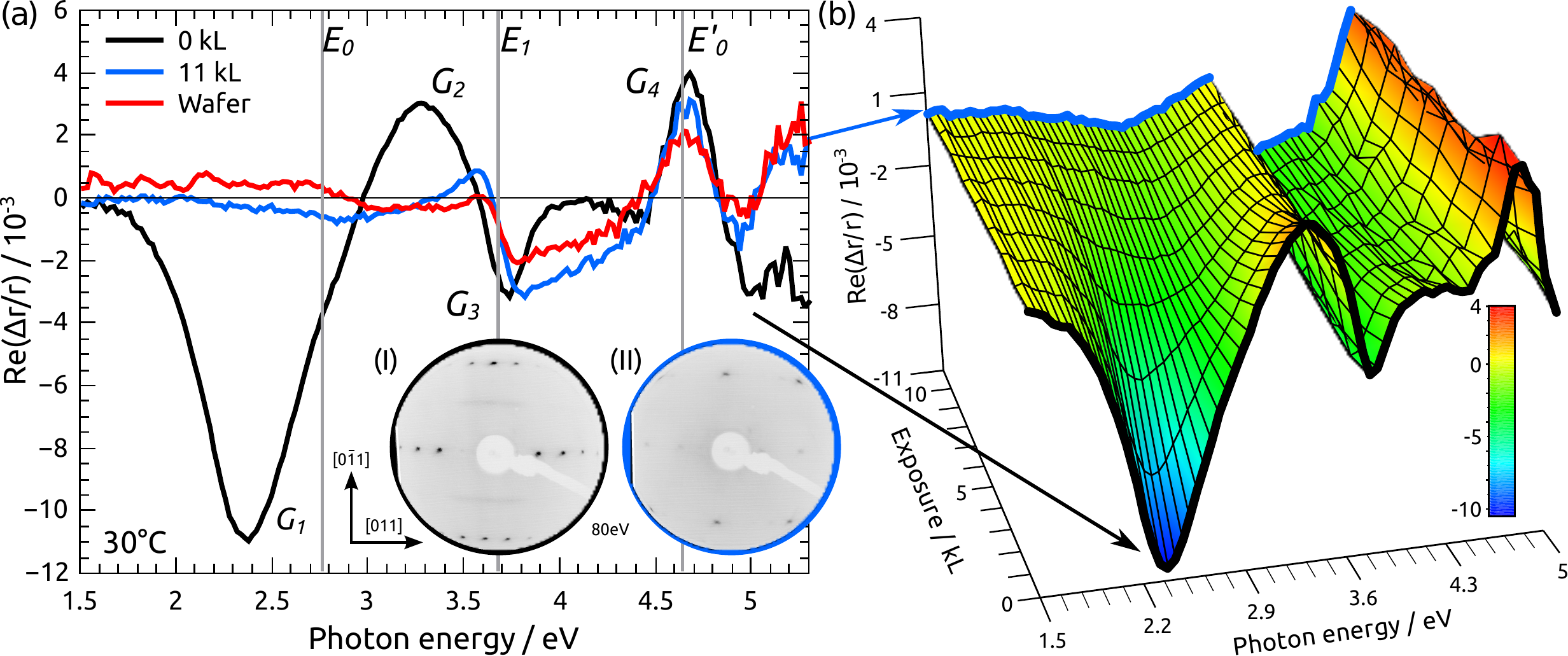}
	\caption{(a) RA spectra of the Ga-rich $(2 \times 4)$ surface before (black) and after (blue) H$_2$O exposure with the critical point energies $E_i$ (grey lines) \cite{Zollner_critical_points_GaP_1993}. The peaks in the RA spectra are labelled $G_i$. The red curve shows the spectrum of an epi-ready GaP(100) wafer before deoxidation. Insets shows LEED images of the clean (I) and exposed (II) surface measured at $E=80$\,eV. (b) Colourplot  showing the evolution of the RA spectrum during H$_2$O exposure.}
	\label{fig:RAS-Ga-reich-LEED}
	\end{center}
\end{figure}

Ga-rich samples were prepared with MOVPE under RAS in-situ control before they were transferred to UHV. At first, the clean surface was characterized by ultraviolet and X-ray photoelectron spectroscopy (UPS/XPS) as well as LEED: The diffraction patterns are typical for a $(2 \times 4)$ surface reconstruction (cf. figure \ref{fig:RAS-Ga-reich-LEED}(a)) \cite{Toeben_RAS_LEED_STM_GaP_2001}. PES confirmed a clean surface free of oxygen and carbon. Afterwards, an RA spectrum of the pristine surface was recorded in the adsorption chamber (figure \ref{fig:RAS-Ga-reich-LEED}(a)), agreeing with the spectrum recorded in MOVPE ambient. The prominent negative anisotropy signal, labelled $G_1$, around 2.4\,eV originates from the Ga-Ga bonds in the [011] direction, the maximum $G_2$ from transitions between anion-dimer states and surface resonances, and the higher-energetic features $G_3$, $G_4$ around the critical points $E_1$ and $E_0^|$ \cite{Zollner_critical_points_GaP_1993} arise from surface-modified bulk transitions \cite{Frisch_Ga-rich_GaP_1999}.

Figure \ref{fig:RAS-Ga-reich-LEED}(b) shows colour-coded, continuously measured RA spectra during exposure to H$_2$O. A gradual suppression of the spectral features and partly a shift in energy can be observed with increasing dosage. The maximum exposure was defined by a saturation of changes in the RA spectrum, which was 11\,kL in this case. A spectrum measured with higher resolution after exposure (figure \ref{fig:RAS-Ga-reich-LEED}(a), blue line) reveals that the negative anisotropy $G_1$, typical for the surface Ga-Ga bonds of the surface reconstruction, is largely suppressed and has transformed into a weak, broad minimum around 2.8\,eV. The negative peak $G_3$ near $E_1$ has shifted $\sim80$\,meV to higher energies and at $E_0^|$, the peak $G_4$ was conserved.

\begin{figure}[h]
	\begin{center}
	\includegraphics[width=\linewidth]{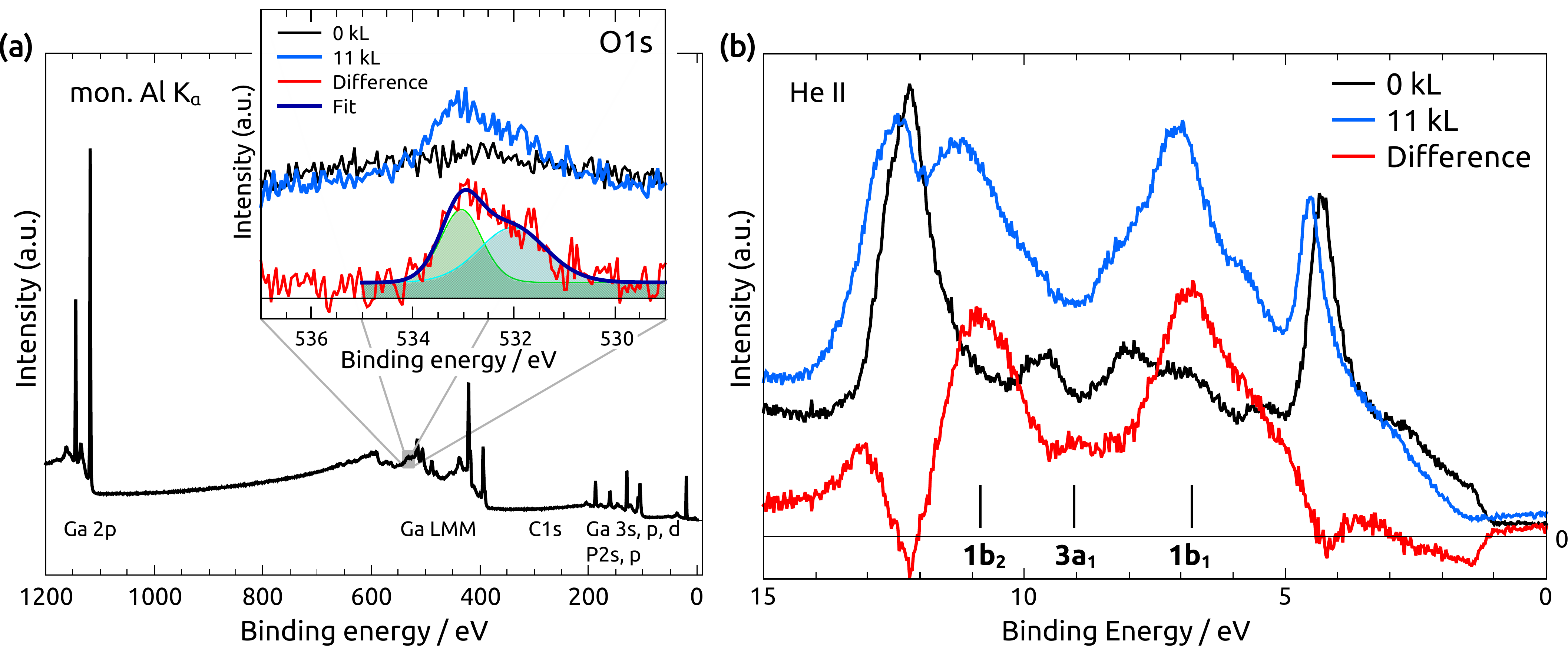}
	\caption{(a) XPS overview spectrum of the  clean Ga-rich surface. The inset shows the region around the O1s line before and after exposure, as well as a deconvoluted difference spectrum. (b) He II spectrum of the Ga-rich surface before and after exposure to H$_2$O with a difference spectrum and the positions of the orbitals of molecular H$_2$O, 1b$_{1,2}$ and 3a$_1$ \cite{Henderson2002}.}
	\label{fig:XPS-UPS-Ga-reich}
	\end{center}
\end{figure}

X-ray photoelectron spectra of the O1s line are plotted in figure \ref{fig:XPS-UPS-Ga-reich}(a). The take-off angle of the photoelectrons was 60$^\circ$ against normal emission to increase surface sensitivity. An overview spectrum shows no signal of carbon or oxygen contamination of the unexposed surface. After exposure, a weak oxygen signature containing two components can be detected. Since the background involves a Ga LMM Auger line for the monochromated Al K$_\alpha$ X-ray source, a difference spectrum was used for further analysis. The intensity of the two spectra was normalized, the spectrum after exposure was shifted in energy to account for the shift of the whole spectrum (see paragraph below) and finally, the first spectrum was subtracted. A deconvolution revealed two contributions of the signal centred at 533.0 and 532.0\,eV. We could, however, not  detect any changes of the phosphorous or gallium emission lines. A quantitative analysis applying the inelastic mean-free paths of GaP and H$_2$O \cite{Gergely_el_mean_free_path_GaP_2000, Akkerman_electron_inelastic_interaction_water_1999} results in a coverage in the order of 0.25($\pm$0.2) monolayers of oxygen on the surface. LEED patterns of the exposed surface (inset of figure \ref{fig:RAS-Ga-reich-LEED}) mainly exhibit the signature of the $(1 \times 1)$ bulk with very weak residual spots, probably stemming from the original $(2 \times 4)$ reconstruction.

He II valence band spectra of the pristine and the exposed surface are given in figure \ref{fig:XPS-UPS-Ga-reich}(b). In general, significant features of the valence band are conserved, with major additional contributions between 5 and 12\,eV binding energy. The difference spectrum was calculated in analogy to the XP spectrum: Intensities were normalized (to the first prominent peak), the spectrum of the surface with 11\,kL H$_2$O shifted in energy to match the unexposed spectrum and finally the latter was subtracted. In this difference spectrum, we observe two main peaks at the binding energies 6.8 and 10.9\,eV, a small peak at 9.1\,eV, and a shoulder at 5.3\,eV. Furthermore, a reduction of the emission near the valence band maximum can be observed. Binding energies $E_B$ of the spectrum were shifted by $\Delta E_B\approx250\,$meV to higher binding energies after exposure, while the secondary electron cut-off, $E_{SC}$, was shifted by $\Delta E_{SC}\approx170\,$meV (not shown here).

\begin{figure}[h]
	\begin{center}
	\includegraphics[width=.95\linewidth]{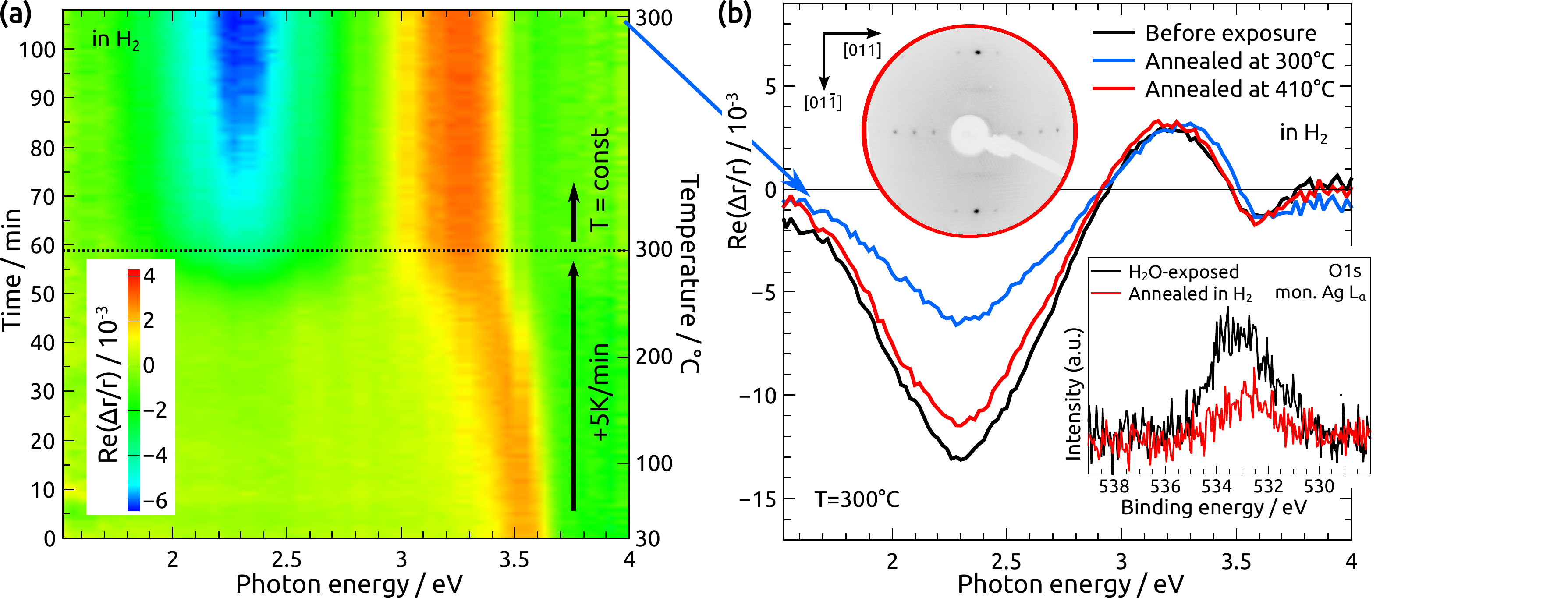}
	\caption{(a) RA colourplot (top view) while annealing the exposed surface in H$_2$ ambient at a temperature ramp of 5\,K/min. The spectrum at $t=0$ corresponds to the blue spectrum in Fig. \ref{fig:RAS-Ga-reich-LEED}. After reaching 300$^\circ$C, the temperature was held constant. (b) RA spectra directly after preparation, after annealing the water-exposed surface in hydrogen to 300$^\circ$C, and finally after annealing to 410$^\circ$C, measured at 300$^\circ$C, respectively. Insets show XP spectra of the O1s line after water exposure and after annealing in hydrogen as well as a LEED image after annealing at $E=60$\,eV.}
	\label{fig:Abheizrun-Ga-reich}
	\end{center}
\end{figure}

To obtain information about the reversibility of the H$_2$O adsorption, samples were transferred back to MOVPE to anneal them with RAS monitoring. Figure \ref{fig:Abheizrun-Ga-reich}(a) shows a colourplot of a Ga-rich sample, which was annealed in H$_2$ atmosphere without precursor supply applying a temperature ramp of 5\,K/min from room temperature up to 300$^\circ$C. As the spectra were still indicating a change of the surface, temperature was held constant for another 50\,min until we could not observe any more changes. Afterwards, a spectrum with higher resolution was acquired (figure \ref{fig:Abheizrun-Ga-reich}(b), blue line) and the sample was annealed further to the temperature typical for the removal of excess phosphorous after growth, 410$^\circ$C. At this temperature, precursor fragments desorb from the surface, leaving the P dimers intact. A preferential desorption of P from the surface would require temperatures $> 470^\circ$C \cite{Toeben_RAS_LEED_STM_GaP_2001}. Back at 300$^\circ$C, a spectrum for comparison to the initial MOVPE-prepared Ga-rich surface was recorded. The negative anisotropy in the low-energetic region typical for the surface reconstruction could almost be restored as well as the LEED patterns (inset of figure \ref{fig:Abheizrun-Ga-reich}(b)). The higher-energy features of the surface-modified bulk transitions are fully and more quickly restored. After annealing to 410$^\circ$C, we still find a very weak oxygen signal at $E_B=532.7$\,eV with roughly a third of the integrated signal intensity compared to the exposed surface, see inset of figure \ref{fig:Abheizrun-Ga-reich}(b).

\subsection{P-rich, buckled dimer surface reconstruction}
\label{sec:P-rich}

\begin{figure}[h]
	\begin{center}
	\includegraphics[width=.95\linewidth]{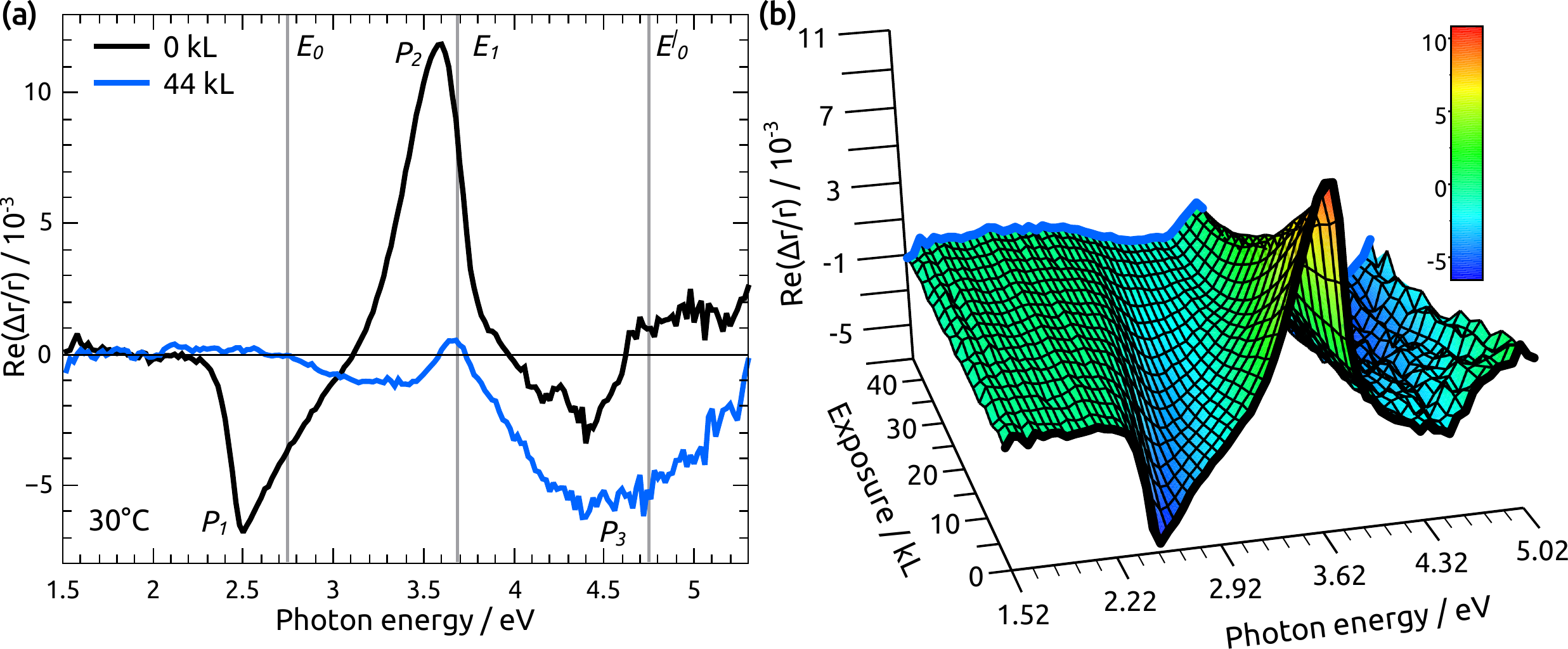}
	\caption{(a) RA spectra of the P-rich surface before (black) and after (blue) H$_2$O exposure with the critical points $E_i$ and peak labels $P_i$. (b) Colourplot acquired during exposure.}
	\label{fig:RAS-P-reich}
	\end{center}
\end{figure}

The P-rich surface reconstruction features a sequence of a negative anisotropy $P_1$ around 2.5\,eV and a strong positive peak $P_2$ around $E_1$. The former signature is most characteristic for this surface geometry and its lower-energy part originates from the hydrogen termination of the P dimers \cite{Hahn_P-rich_GaP_calculated_2003, Schmidt_InP_hydrogen_stabilized_2003}, see figure \ref{fig:RAS-P-reich}(a). $P_2$ stems from a surface modified bulk transition, according to the observations made on the corresponding RAS signal of P-rich InP(100) \cite{Visbeck_temperature_dependence_RAS_InP_2001}. Similarly to the Ga-rich surface, water exposure leads to a suppression of most RA signatures, yet the required exposure leading to a saturation is four times higher (cf. figure \ref{fig:RAS-P-reich}(b)). After exposure, we observe a broad and weak minimum between the critical points $E_0$ and $E_1$, similar to the Ga-rich surface after exposure. In the higher energetic part, however, the spectrum exhibits a second, broad and intense minimum, $P_3$, around $E_0^|$.

\begin{figure}[h]
	\begin{center}
	\includegraphics[width=.75\linewidth]{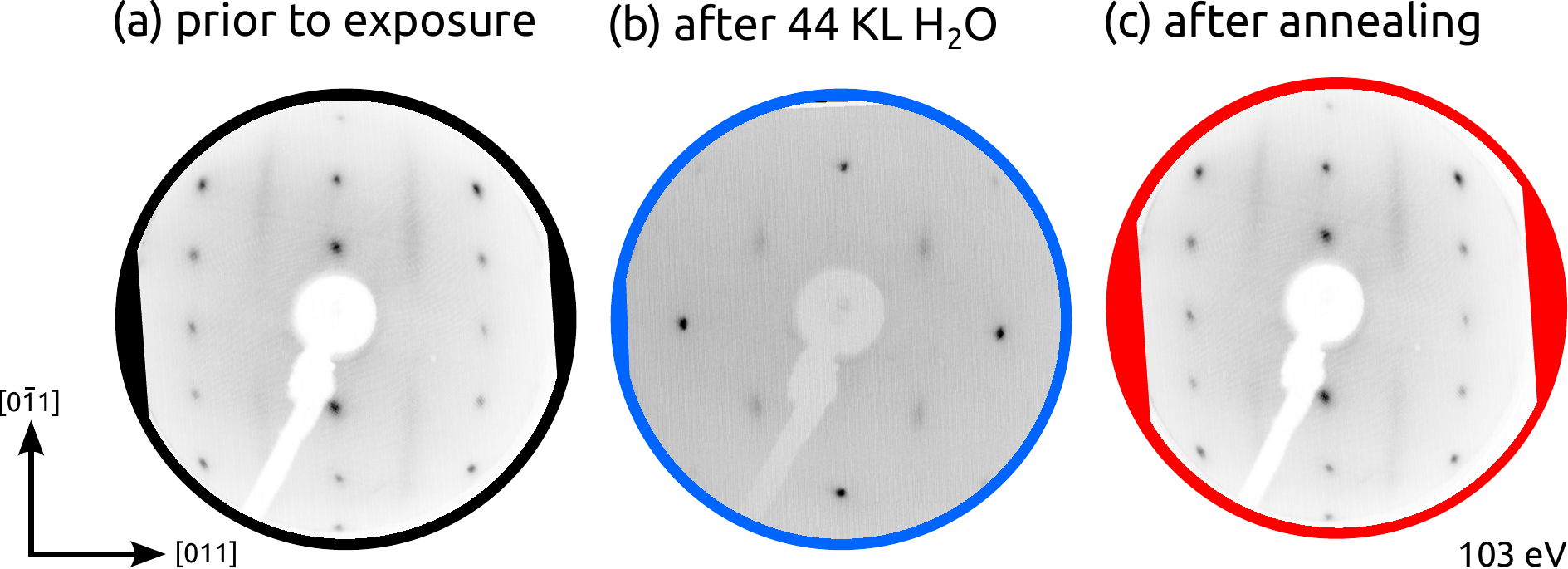}
	\caption{LEED images of the clean (a) and the exposed (b) and the nitrogen-annealed (c) P-rich surface at $E=103$\,eV.}
	\label{fig:LEED-P-rich}
	\end{center}
\end{figure}

\begin{figure}[h]
	\begin{center}
	\includegraphics[width=\linewidth]{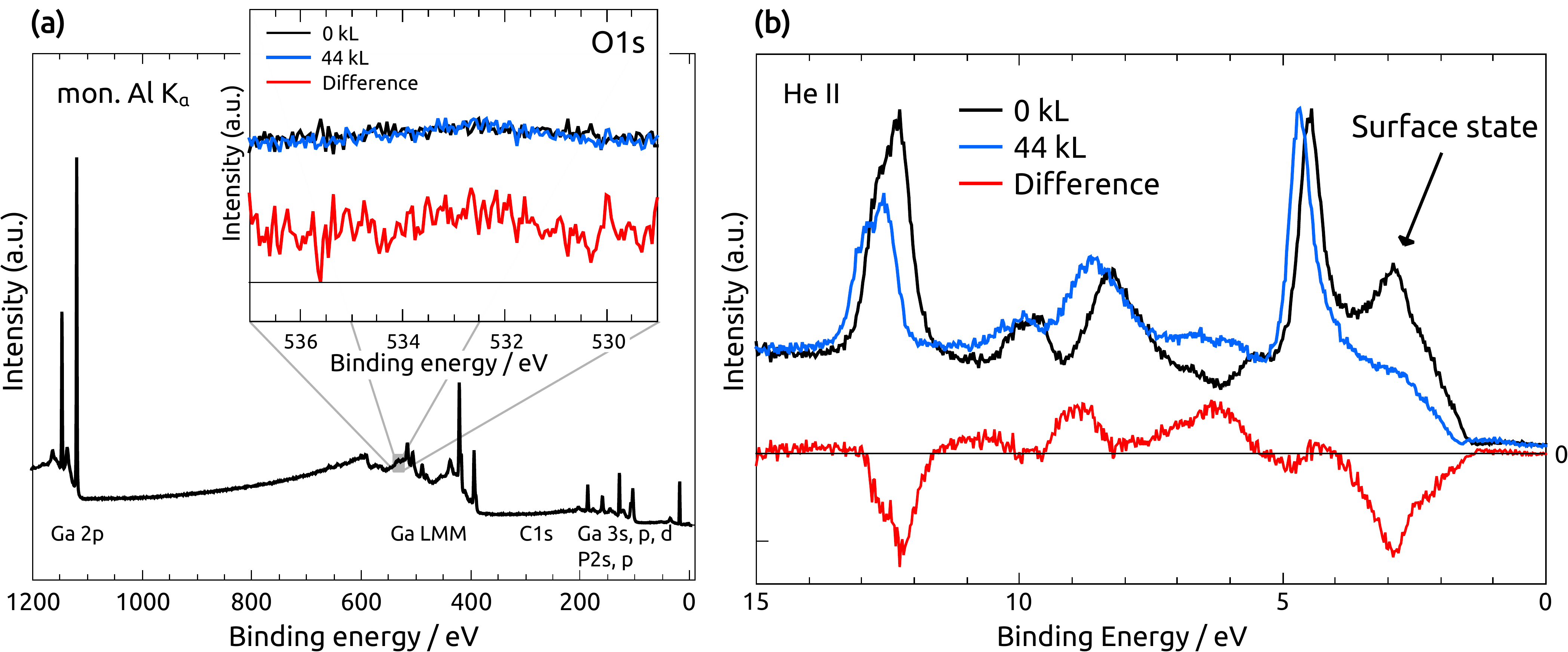}
	\caption{(a) XPS overview spectrum of the  clean P-rich surface. The inset shows the region around the O1s line before and after exposure, as well as the difference spectrum. (b) He II spectrum of the P-rich surface before and after exposure to H$_2$O with difference spectrum. }
	\label{fig:XPS-UPS-P-reich}
	\end{center}
\end{figure}

Before water exposure, LEED patterns display spots at half-order, typical for the  $p(2 \times 2)/c(4 \times 2)$ surface reconstruction (figure \ref{fig:LEED-P-rich}(a)). The streaks at half order along $[0\bar{1}1]$ originate from a mixture of $p(2 \times 2)$ and $c(4 \times 2)$ domains \cite{Li_InP_surface_1999}. XPS -- again tilted 60$^\circ$ -- does not reveal any clear signature of oxygen (cf. figure \ref{fig:XPS-UPS-P-reich}(a)). In contrast, the LEED features change significantly upon exposure: Instead of the  $p(2 \times 2)/c(4 \times 2)$ reconstruction, a $(1 \times 1)$ structure with an additional, blurry $c(2 \times 2)$ superstructure can be observed (figure \ref{fig:LEED-P-rich}(b)). 

\begin{figure}[h]
	\begin{center}
	\includegraphics[width=.95\linewidth]{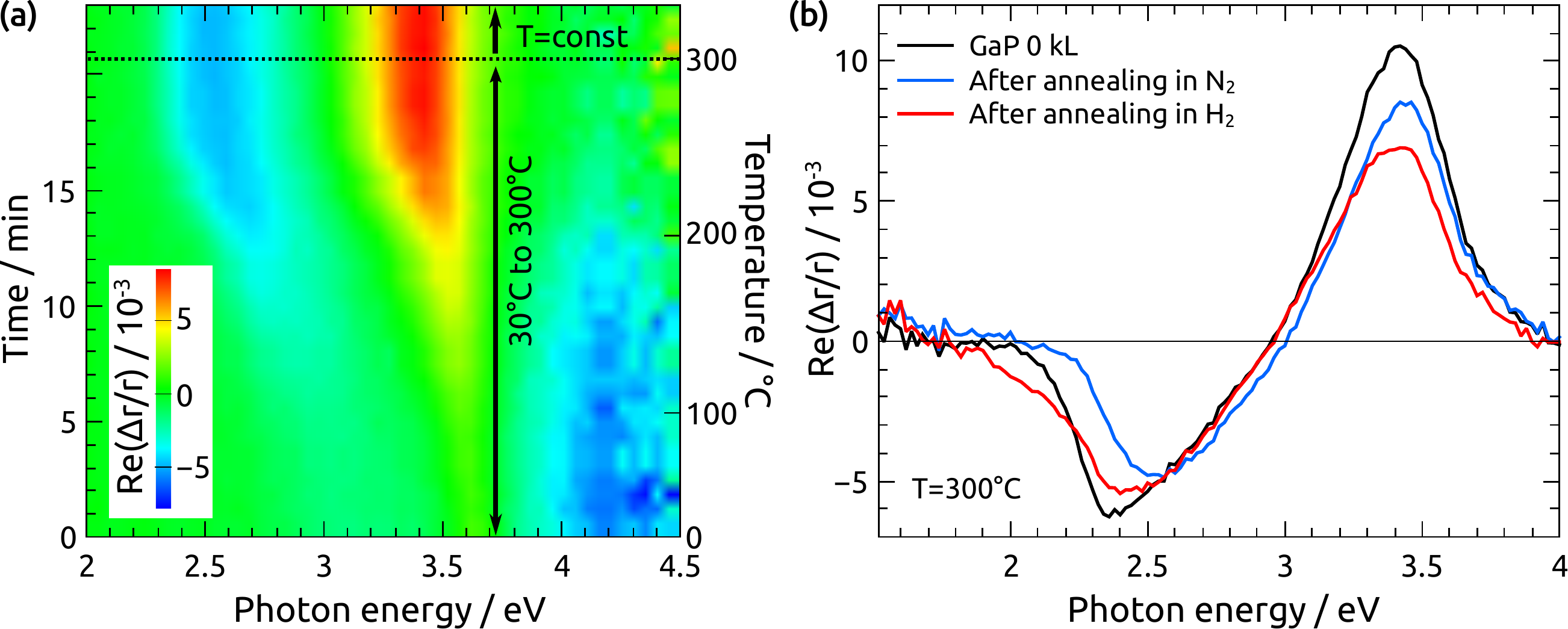}
	\caption{(a) Colourplot (top view) while annealing the exposed surface in N$_2$ ambient at a temperature ramp of 18\,K/min. The spectrum at $t=0$ corresponds to the blue spectrum in Fig. \ref{fig:RAS-P-reich}. After reaching 300$^\circ$C, the temperature was held constant. (b) RA spectra at 300$^\circ$C directly after preparation and after specific annealing steps.}
	\label{fig:Abheizrun-P-reich}
	\end{center}
\end{figure}

Figure \ref{fig:XPS-UPS-P-reich}(b) shows He II UP spectra before and after exposure. For the difference spectrum, the spectrum of the exposed surface had to be shifted again in energy. Unlike for the Ga-rich surface, we cannot observe a clear signature of H$_2$O here. Four very weak contributions can be found at the energies 6.1, 6.9, 8.9 and 10.9\,eV. The characteristic peak associated to the surface state of the P-rich reconstruction \cite{ChemPhysChem_III-V_2012} around 2.9\,eV has disappeared, which would be expected for a modification of the surface reconstruction. Binding energies were shifted by $\Delta E_B\approx180\,$meV and the secondary electron cut-off by $\Delta E_{SC}\approx240\,$meV to higher binding energies after exposure. 

Samples were again transferred back to VPE ambient employing the UHV transfer chamber. Annealing in purified nitrogen allows for a recovery of most of the initial RA spectrum, setting in at temperatures between 200 and 250$^\circ$C as displayed in figure \ref{fig:Abheizrun-P-reich}(a). At the same time, the high-energetic peak $P_3$ between 4 and 4.5\,eV disappears. Figure \ref{fig:Abheizrun-P-reich}(b) compares spectra at 300$^\circ$C after specific annealing steps. Most of the spectrum could be recovered after annealing up to 300$^\circ$C in N$_2$. The low-energetic part of the negative anisotropy around 2.5\,eV, however, fails to be recovered in nitrogen even at temperatures of 410$^\circ$C. By contrast, it can be -- mostly -- restored upon supply of H$_2$ at this temperature. The LEED patterns of the  $p(2 \times 2)/c(4 \times 2)$ reconstruction, including the half-order streaks, are regenerated after annealing in either process gas, see figure \ref{fig:LEED-P-rich}(c).

\section{Discussion}

\subsection{Ga-rich, mixed dimer surface reconstruction}

The Ga-rich, $(2 \times 4)$ surface reconstruction is clearly more reactive than the P-rich,  $p(2 \times 2)/c(4 \times 2)$ surface reconstruction as indicated by the RA spectra during exposure and the resulting coverage observed with PES. The RA signatures related directly to the specific surface reconstruction, i.e. the peaks $G_1$ and $G_2$, as well as the LEED patterns of the $(2 \times 4)$ reconstruction mostly disappear (cf. figure \ref{fig:RAS-Ga-reich-LEED}). This diminished signal can in principle be caused by the establishment of an isotropic adsorbate layer suppressing the RAS (LEED) signal or the breaking of chemical bonds initiated for example by the formation of Ga-O-Ga or Ga-[OH]-Ga bonds. The latter scenario, in form of a modification of the Ga-Ga bonds in [011] direction, is supported by the finding that the RAS peak $G_1$, which is related to these bonds \cite{Frisch_Ga-rich_GaP_1999}, disappears most quickly. This would also conform to the models proposed for this surface \cite{Wood_GaP_InP_2012, Jeon_GaP_2012} and in case of the closely related InP, the lower-energetic feature is indeed most sensitive to surface chemistry \cite{Hannappel_RAS_InP_2000}. A complete loss of surface ordering, however, would lead to a loss of most features of the spectrum, as observed e.g. for the oxidation of the In-rich InP(100) surface \cite{Chen_InP_oxidation_RAS_2002}, which is clearly not observed here. The weak negative anisotropy around 3\,eV, observed after exposure, could be a remainder of the original anisotropy $G_1$. Very weak residual LEED spots support this view. No new superstructure is visible, which would suggest that the remaining RAS features are surface-modified bulk transitions or another anisotropic modification not visible to LEED.

Above 4.5\,eV, the RA spectrum follows quite closely the signal of an oxidized, ``epi-ready'' wafer (figure \ref{fig:RAS-Ga-reich-LEED}(a)). Especially $G_4$ is almost conserved, both in position and magnitude, emphasizing its nature as a surface-modified bulk transition. A clear shift of $G_3$ in the region around $E_1$ can be observed, which could be a manifestation of the linear electro-optic effect \cite{Leo_PRB_1989} due to the dipole at the surface, that also causes the shift of PE spectra, or doping induced by chemisorbed species. Between 3.7 and 4.5\,eV, the $(2 \times 4)$ surface exhibits a negative anisotropy previously absent, but also present in a similar form for the wafer. This indicates that there exists some similarity to the oxide, the increased magnitude of this feature (as well as $G_4$) for the exposed surface could be explained by a higher grade of ordering or a reduced roughness of the MOVPE-prepared surface when compared to the wafer.  

X-ray photoelectron spectroscopy confirms the presence of two species of oxygen on the exposed Ga-rich surface. The contribution with the higher binding energy at 533\,eV stems probably from molecular H$_2$O \cite{Henderson2002}, while we ascribe the weaker peak at 532\,eV to surface hydroxyl species. The weakness of the signal (see figure \ref{fig:XPS-UPS-Ga-reich}(a)), however, indicates that the coverage is in the order of 0.25($\pm0.2$) monolayers (2 atoms per unit cell of the surface reconstruction), which is in line with the absence of any detectable chemical shift of the P or Ga lines within the given surface sensitivity of our setup. Effective coverages in the literature were significantly higher, as either low-temperature adsorption was applied \cite{Henrion_water_InP_2000, Chung_H2O_GaAs_1998}, resulting in a high sticking coefficient, or very high dosages in the order of $10^{9}-10^{10}$\,L \cite{webb_H2O_GaAs_1982}. The O1s emission becomes significantly stronger upon tilting the sample, which indicates that the adsorption happens at the very surface and that oxygen is probably not diffusing into the bulk creating Ga-O-P bonds as suggested in one scenario of Wood et al. \cite{Wood_GaP_InP_2012}. This is also in line with our finding that an annealing at already 300$^\circ$C can restore most of the RA spectrum, while the deoxidation of an oxidized wafer requires temperatures in the order of 600$^\circ$C \cite{Kleinschmidt_dimer_flipping_2011}.

UPS displays the appearance of four additional contributions superimposed on the valence band structure of the Ga-rich surface. The higher-energetic peaks fit quite well to the valence band structure of molecular water found for similar systems \cite{Henrion_water_InP_2000, Henderson2002}, their absolute energetic positions depending on the semiconductor surface \cite{Henderson2002}. The shoulder at 5.3\,eV could, in analogy to studies of the InP(110) surface by Henrion et al. \cite{Henrion_water_InP_2000}, be ascribed to Ga-OH bonds, which supports the XPS interpretation above. These findings suggest a combination of chemisorbed hydroxyl groups and molecularly (co)adsorbed water on the surface, similar to GaAs surfaces \cite{webb_H2O_GaAs_1982}. The final step of a dehydrogenation of the hydroxyl groups, as found for GaAs(100) surfaces at temperatures in the order of 330-430$^\circ$C \cite{Chung_H2O_GaAs_1998} and which was also proposed for GaP(100) in ref. \cite{Jeon_GaP_2012}, therefore seems to be not the dominant reaction path at room temperature. The coverage in ref. \cite{Jeon_GaP_2012} is kept at 0.125 monolayers (one molecule per surface unit cell), though basically five adsorption sites were identified for molecular H$_2$O, of which two pairs are symmetric. Our rather high exposures result in higher coverages, possibly occupying two of these symmetric sites per unit cell.

A change of band bending, $\Delta eV_{BB}$, shifts both the binding energy, $E_B$, and the secondary electron cut-off, $E_{SC}$ \cite{Jaegermann_modern_asp_elchem}. This can be either a downward band bending or the reduction of an existing upward band bending (neglecting charge-transfer, which would have the same effect). The work function/secondary electron cut-off, on the other hand, can be shifted by a change of the surface dipole, $\Delta \chi_s$, as well as a change in band bending, $\Delta eV_{BB}$:

\begin{equation}
 \Delta E_{SC}=\Delta\Phi=\Delta \chi_s-eV_{BB}
\end{equation}

The observed shifts $\Delta E_B$ and $\Delta E_{SC}$ suggest a downward change of band bending by $\Delta eV_{BB}\approx250\,$meV. The shift $\Delta E_{SC}\approx170\,$meV, however, differs slightly towards a higher work function, pointing to the existence of a positive dipole at the surface after exposure. We assume that this change of the work function originates from the dipole component of water molecules perpendicular to the surface implying an orientation of the molecules with the hydrogen towards the surface. In summary, we presume that the adsorbed water causes a downward change of band-bending and creates a positive dipole on the surface.

After annealing in hydrogen (figure \ref{fig:Abheizrun-Ga-reich}), we still find a very weak oxygen signal, which exhibits about one third of the original intensity and is shifted in energy. This could indeed be the remaining oxygen after dehydrogenation of the hydroxyl groups, without the molecular H$_2$O. Peak $G_1$ of the RA signal can, unlike $G_2$, not be fully restored by annealing up to 410$^\circ$C, which is another evidence for an attack on the Ga-Ga bonds at the surface in [011] direction. The oxygen could be inserted into those bonds, which is thought to be the configuration with the lowest formation energy \cite{Wood_GaP_InP_2012}. LEED shows again the symmetry of the original surface reconstruction, so most of the original surface symmetry is restored.

\subsection{P-rich, buckled dimer surface reconstruction}

The P-rich surface is clearly more stable towards water-induced surface modifications as indicated by the significantly higher H$_2$O dosage required to saturate the RA signal and the lower resulting coverage evidenced by PES. This is probably due to the quite inert P dimer stabilized by an additional hydrogen atom, similar to the hydrophobic properties of hydrogen-terminated Si \cite{Yau_hydrophobic_Si_1995}. Saturation exposures for water adsorption on very inert TiSe$_2$ surfaces \cite{May_PRL_2011} are in the same order of magnitude as for our experiments on the P-rich surface, emphasizing that the surface exhibits a very low reactivity. As for the Ga-rich surface, the RA features $P_1$ and $P_2$, specific for the reconstruction, are greatly reduced. A negative anisotropy between $E_0$ and $E_1$ remains, which is stronger and at higher energies when compared to the Ga-rich surface (figure \ref{fig:RAS-P-reich}(a)). The signal is less structured in the high-energetic region, but exhibits a very intense negative anisotropy $P_3$. This feature is probably related to the $c(2 \times 2)$ superstructure observed in LEED (figure \ref{fig:LEED-P-rich}(b)): The disappearance of the RAS peak $P_3$ during annealing is accompanied by the reappearance of $P_1$, $P_2$ and also restores most of the original LEED signature (figure \ref{fig:LEED-P-rich}(c)). The slightly diffuse, but reproducible $c(2 \times 2)$ LEED diffraction patterns are a distinct feature of the exposed P-rich surface prior to annealing.

With XPS, we cannot detect any oxygen signal after 44\,kL H$_2$O exposure at room temperature (figure \ref{fig:XPS-UPS-P-reich}(a)). This is similar to results for InP(100) surfaces that were exposed to molecular oxygen, where the authors did not find a significant oxygen coverage for exposures in the order of $10^5$\,L at ambient temperature for the P-rich surface, in contrast to the In-rich surface \cite{Chen_InP_oxidation_RAS_2002}. The presence of any carbon can impact the oxygen uptake of the P-rich surface significantly. However, in most cases we could not detect any carbon on the surfaces after exposure. In the few cases, where carbon was detected due to deficient UHV conditions, a relatively strong oxygen signal was indeed observed.

Essential features of the He II UPS valence band spectrum of the P-rich surface are retained after exposure. Apart from the disappearance of the peak around 3\,eV (figure \ref{fig:XPS-UPS-P-reich}(b)), which we tentatively attribute to charge-transfer from a phosphorous-related surface state \cite{Montgomery_H2O_InP_1982}, the three higher-energetic additional peaks could in principle be attributed to molecular water. However, the intensity ratios do not fit well, which could indicate a different state of the adsorbate (see below). Together with XPS analysis, the weak signal suggests a very low coverage (below 0.1 monolayers) of H$_2$O(/OH) on the surface.

\begin{figure}[thbp]
	\begin{center}
	\includegraphics[width=.65\linewidth]{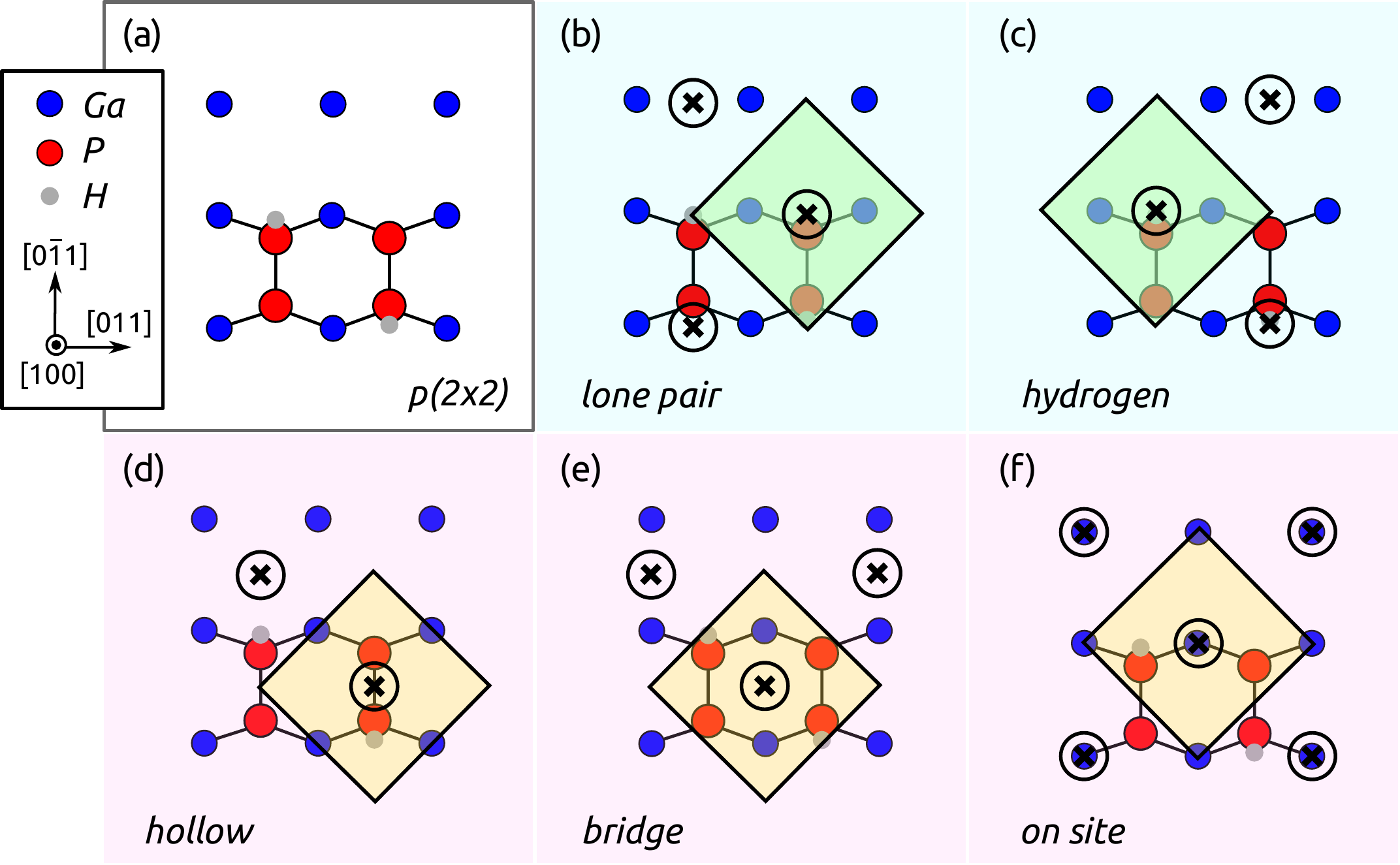}
	\caption{P-rich, $p(2 \times 2)$ surface reconstruction with adsorption spots (+) in different configurations creating a $c(2 \times 2)$ superstructure. The sites (b), (c) and (e) represent different realisations of a bridge configuration.}
	\label{fig:P-rich_c2x2}
	\end{center}
\end{figure}

The $c(2 \times 2)$ superstructure is in principle compatible with three configurations of the adsorbate: a 4-fold hollow, a bridge or an on-top configuration (with respect to the Ga atoms) as depicted in figure \ref{fig:P-rich_c2x2}(d-f) and which is also a scenario taken into account by Wood et al., albeit for a relaxed $(2 \times 2)$ surface \cite{Wood_GaP_InP_2012}. If every adsorption spot was occupied, this would result in a coverage of half a monolayer. Taking into account the findings of PES quantification suggests that only a small fraction of the spots is occupied.

An analysis of the shifts $\Delta E_B$ and $\Delta E_{SC}$ suggest again a change in band bending, either downward band bending or reduction of an existing upward band bending. The value $\Delta E_B\approx180\,$meV, however, is only slightly smaller than for the Ga-rich surface, though the coverage is significantly lower. The work function is reduced more than would be expected for a change in band bending. The dipole change on the surface $\Delta E_{SC,\chi}$ is therefore negative, suggesting an orientation of the water molecule with its oxygen end towards the surface.

\begin{figure}[thbp]
	\begin{center}
	\includegraphics[width=.5\linewidth]{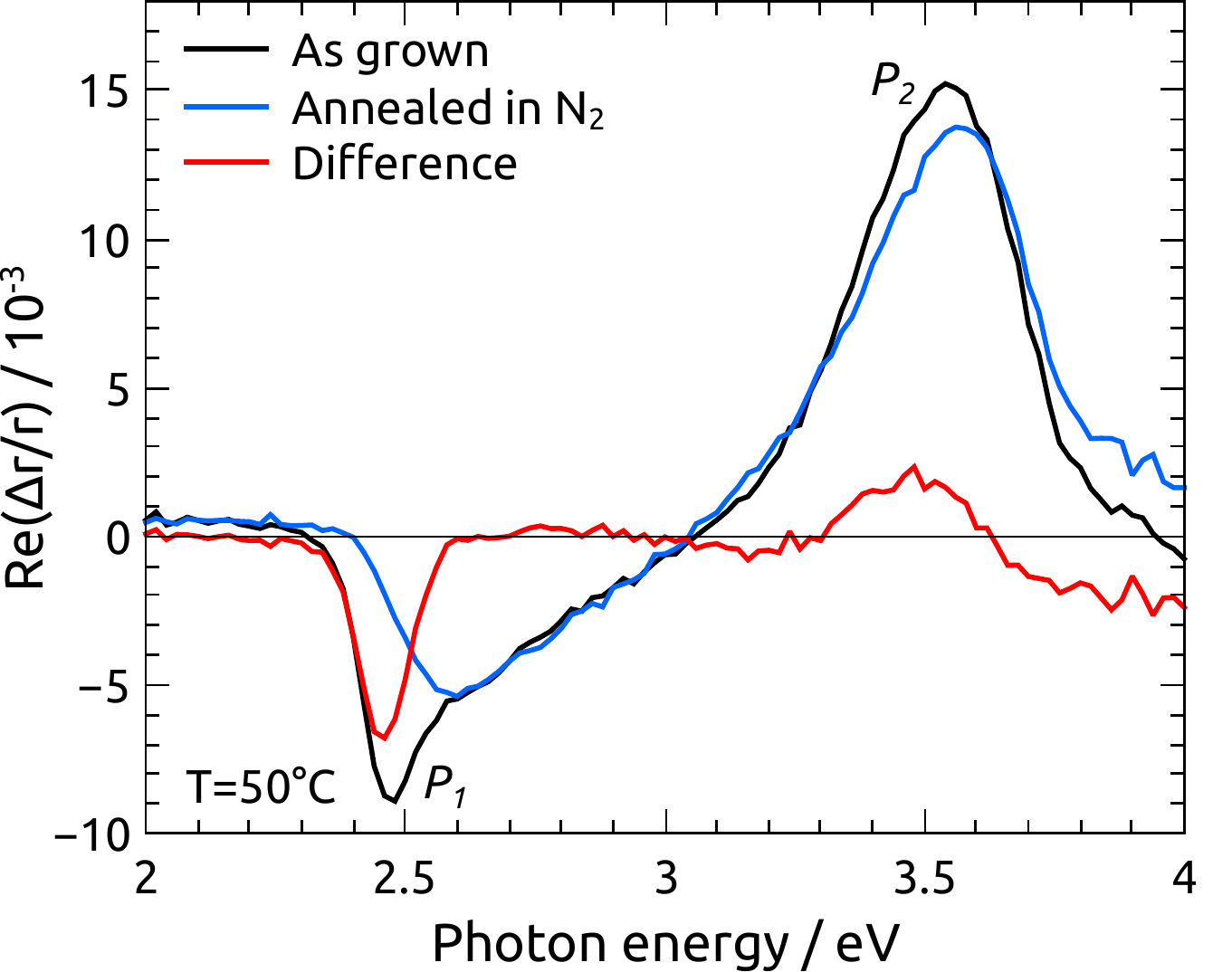}
	\caption{RA spectra in the reactor of the freshly grown sample and after annealing of the water-exposed surface in nitrogen.}
	\label{fig:RAS-diff-annealed}
	\end{center}
\end{figure}

Annealing the surface in hydrogen can recover the LEED patterns, including the half-ordered streaks, and also largely the RA spectrum. The latter is recovered in the general features, but lacks some intensity. This suggests that some ordering of the surface is lost during annealing. Interestingly, the finding that $P_3$ disappears simultaneously with the reappearance of $P_1$ and $P_2$ suggests that this feature is related to the adsorbate causing the $c(2 \times 2)$ superstructure. The temperature associated with this desorption, 200 to 250$^\circ$C, shows that the bonding is not very strong. The same annealing procedure in nitrogen process gas recovers most of $P_2$ and the LEED patterns as well. The low-energetic part of $P_1$, however, is not restored at all, as shown in the difference spectrum in figure \ref{fig:RAS-diff-annealed}. This is exactly the feature that becomes very intense at low-temperature RAS \cite{Hannappel_RAS_InP_20K_2000} and is related to the hydrogen atom at the P dimer \cite{Schmidt_InP_hydrogen_stabilized_2003}. A likely interpretation is that the annealing in nitrogen leaves an intact, buckled P dimer, which is not hydrogen-stabilized. The P-H bond can only (partly) be restored upon the exposure to hydrogen in VPE ambient. This would not contradict the restored half-ordered streaks of the  $p(2 \times 2)/c(4 \times 2)$ LEED signature: before the discovery of the involvement of hydrogen in the surface reconstruction, it was explained by naked, buckled P dimers \cite{Li_InP_surface_1999}. Annealing of pristine surfaces with intact P-H bond in nitrogen at these temperatures does, however, not change the RA signal. This means, that the bonding of the hydrogen atoms is weakened by the adsorbate and that they desorb together with the adsorbate or are already removed during the adsorption process at room temperature.

In summary, we assume that only the topmost atomic layer, not involving Ga atoms, participates in the reaction during exposure of the P-rich surface. In the following, we will discuss possible reactions paths based on the findings outlined above that could lead to the state of the P-rich surface observed after exposure. \textsc{I.} A first path could be the dissociative adsorption of water resulting in one P-H and one P-OH bond per dimer. The desorption during annealing in form of H$_2$O would leave a P dimer without H. This does, however, not fit to the molecular water-like UPS signature. \textsc{II.} An alternative reaction path would involve molecular water that creates a hydrogen bond with the lone electron pair of the P-atom, which is buckled upwards, either in an on-top or a bridge configuration, see figure \ref{fig:P-rich_c2x2}. The hydrogen bond could then weaken the original P-H bond leading to a desorption of molecular H$_2$ already during exposure. This mechanism would, however, result in a positive dipole on the surface contrasting our findings above. \textsc{III.} A third adsorption path would be a hydrogen bond between the oxygen atom of the water molecule and the hydrogen atom of the P dimer, in agreement with the negative dipole we found for the P-rich surface. The latter two mechanisms could also explain the $c(2 \times 2)$ superstructure. For possible bridge configurations, one would also have to take into account the possibility of dimer-flipping \cite{Kleinschmidt_dimer_flipping_2011} at room temperature, where one P dimer could flip to increase the interaction with the H$_2$O molecule. \textsc{IV.} Finally, molecular water could form a kind of coordination compound with the P dimer, weakening the original P-H bond. This fourth path would fit best to our findings.

\section{Summary and conclusion}

We have investigated the interaction of adsorbed water with the Ga-rich, $(2 \times 4)$ and the P-rich,  $p(2 \times 2)/c(4 \times 2)$ reconstructions of GaP(100) surfaces prepared by metal-organic vapour phase epitaxy. The behaviour of the surfaces was found to differ significantly, both in reactivity and reaction path. The experimental results found for the Ga-rich surface are similar to other III-V semiconductors \cite{webb_H2O_GaAs_1982,Henrion_water_InP_2000}, with our findings pointing towards a mixture of dissociatively and molecularly adsorbed water. The saturation coverage, defined via the RA spectrum, is in the order of 0.25\,monolayers. P-rich surfaces, on the other hand, exhibit an even lower adsorbate coverage forming a $c(2 \times 2)$ superstructure on the surface. Observations with reflection anisotropy spectroscopy during the annealing of exposed P-rich surfaces suggest a removal of the hydrogen atom of the P-dimer induced by the adsorbate during exposure or a significant weakening of the P-H bond resulting in facilitated desorption. Only the Ga-rich surface exhibits a partial oxidation after annealing and both surfaces can be largely restored without supplying gallium or phosphorous precursors. These observations show that RAS is a highly sensitive in-situ tool for the monitoring of semiconductor surfaces modified by adsorbates and potentially also in liquid environments.

Our findings could also benefit the design of GaP-based photoelectrochemical water splitting devices, as band bending and surface dipoles impact charge separation as well as the charge transfer rate for reduction (or oxidation) of water. Due to the position of valence- and conduction band relative to the water oxidation and reduction potentials, GaP, in a single junction device, can only be used for the reduction of water. The conduction band, however, is located substantially above the water reduction potential, resulting in energy losses \cite{Kaiser2012}. A downward band bending, which is beneficial for the transfer of electrons to the aqueous electrolyte, is found for both surface reconstructions. A positive dipole, increasing the electron affinity, is observed, however, for the Ga-rich surface reconstruction. As this results in a downward movement of the band edges, the Ga-rich surface could therefore be more suitable for water splitting in single junction structures. Also in tandem applications with GaP as photocathode, this shift would be beneficial because the offsets are large enough to enable the reaction at a sufficient rate, but not too large causing excessive energetic losses. Hence the observed increase of the electron affinity of GaP on the Ga-rich surface would reduce the large offset of the conduction band, benefiting a photocathode application for hydrogen evolution. A photoanode application would profit from an increase of the originally small valence band offset with respect to the water oxidation potential and from the fact that water is partially dissociatively adsorbed, facilitating oxidation.

Future work will apply these findings for the development of GaP-based tandem structures for light-induced water splitting with surface functionalization for stability and efficiency using protective layers and electrocatalysts.

\ack
The authors would like to thank Abdelkrim Chemseddine for valuable discussions and Wolf-Dietrich Zabka for experimental assistance. MM May acknowledges a scholarship from Studienstiftung des deutschen Volkes. Part of this work was supported by the German Research Foundation (DFG, Project No. HA 3096/4).

\section*{References}

\providecommand{\newblock}{}

\end{document}